\documentclass[aps,prl,twocolumn,showpacs, superscriptaddress]{revtex4-1}
\usepackage{amsfonts}
\usepackage{bm}
\usepackage{amsmath}
\usepackage{graphicx}
\usepackage{color}
\usepackage[usenames,dvipsnames]{xcolor}
\usepackage{amsmath}
\usepackage{verbatim}

\newcommand{\nc}{\newcommand}
\nc{\eq}{\begin{equation}}
\nc{\eeq}{\end{equation}}
\nc{\eqa}{\begin{eqnarray}}
\nc{\eeqa}{\end{eqnarray}}
\nc{\ar}{\begin{array}}
\nc{\ear}{\end{array}}
\nc{\bfig}{\begin{figure}}
\nc{\efig}{\end{figure}}
\nc{\dg}{\dagger}
\nc{\eps}{\frac{\epsilon}{2}}
\nc{\juuri}{\sqrt{\Omega^2+(\eps)^2}}
\nc{\sx}{\sigma_x}
\nc{\sy}{\sigma_y}
\nc{\sz}{\sigma_z}
\nc{\spl}{\sigma_+}
\nc{\sm}{\sigma_-}
\nc{\Sx}{\bar{\sigma}_x}
\nc{\Sy}{\bar{\sigma}_y}
\nc{\Sz}{\bar{\sigma}_z}
\nc{\Spl}{\bar{\sigma}_+}
\nc{\Sm}{\bar{\sigma}_-}
\nc{\nn}{\nonumber}
\nc{\noi}{\noindent}
\nc{\omt}{\tilde{\omega}}
\nc{\Somt}{S(\omt)}
\nc{\Somtd}{S^{\dg}(\omt)}
\nc{\got}{\gamma_{\omega}(t)}
\nc{\gmot}{\gamma_{-\omega}(t)}
\nc{\po}{\mathcal{P}}
\nc{\qo}{\mathcal{Q}}
\nc{\adg}{a^{\dg}}
\nc{\gammat}{\tilde{\gamma}}
\nc{\Q}{$\mathcal{Q }$}
\nc{\C}{$\mathcal{C }$}
\nc{\kvec}{\mathbf{k}}

\def\bra#1{\mathinner{\langle{#1}|}}
\def\ket#1{\mathinner{|{#1}\rangle}}

\begin{document}

\title{Problem of coherent control in non-Markovian open quantum systems}

\author{Carole Addis}
\affiliation{SUPA, EPS/Physics, Heriot-Watt University, Edinburgh, EH14 4AS, UK}
\email{ca99@hw.ac.uk}
\author{Elsi-Mari Laine} 
\affiliation{Turku Center for Quantum Physics, Department of Physics and Astronomy, University of Turku, FIN-20014 Turku, Finland}
\author{Clemens Gneiting} 
\affiliation{Physikalisches Institut, Albert-Ludwigs-Universit\"{a}t Freiburg, Hermann-Herder-Str.~3, D-79104 Freiburg, Germany}
\author{Sabrina Maniscalco} 
\affiliation{Turku Center for Quantum Physics, Department of Physics and Astronomy, University of Turku, FIN-20014 Turku, Finland}
\affiliation{Center for Quantum Engineering, Department of Applied Physics, Aalto University School of Science, P.O. Box 11000, FIN-00076 Aalto, Finland}
\email[]{smanis@utu.fi} 
\date{\today}

\begin{abstract}
{We critically evaluate the most widespread assumption in the theoretical description of coherent control strategies for open quantum systems. We show that, for non-Markovian open systems dynamics, this fixed-dissipator assumption leads to a serious pitfall generally causing difficulties in the effective modeling of the controlled system. We show that at present, to avoid these problems, a full microscopic description of the controlled system in the presence of noise may often be necessary. We illustrate our findings with a paradigmatic example.}
%
%
\end{abstract}


\maketitle
Motivated by the tremendous progress in quantum technologies, considerable effort has been devoted to minimizing environment-induced decoherence effects. Efficient schemes for compensating harmful noise in quantum systems have been developed utilizing the quantum Zeno effect \cite{QZeno1,QZeno2}, dynamical decoupling strategies \cite{DD1}-\cite{DD3} and optimal control \cite{Clemens}, \cite{OCStart}-\cite{OC5}. Generally, the theoretical description of these techniques in the presence of noise is a daunting task, therefore they are typically studied under a number of assumptions concerning the type of environments the system is interacting with as well as the typical time-scales. Specifically, optimal control techniques have been so far studied, almost exclusively, in the so-called Markovian limit, that is whenever the system-environment interaction is weak and the correlations short living. In this case the master equations describing the open system dynamics are found phenomenologically or derived with microscopic approaches using numerous approximations \cite{Clemens}, \cite{OCStart}-\cite{OC5}. 


Lately, non-Markovian open quantum systems have been drawing a great deal of attention due to their important role in many realistic experimental scenarios \cite{bp}-\cite{Send}. Indeed, when the typical approximations used in the microscopic approaches are not valid, exact approaches are needed to properly describe strong and long-lasting memory effects. 
Further, non-Markovian open quantum systems can be characterized via their capability to gain back information previously lost due to decoherence. Thus, it has been speculated whether the memory effects could be utilized as a resource for quantum information tasks by means of reservoir engineering techniques \cite{R1}-\cite{Rend}. 
Since access and control of the environment is often limited, it is still unclear to what extent such an approach can be carried forward.
In this context, recently many have wondered whether memory effects combined with external control techniques offer a possibility to design an overall superior technique to combat decoherence \cite{DDPaper}, \cite{Daniel}-\cite{Susanna}. Unfortunately, contrary to intuitive reasoning, non-Markovianity is not trivially a resource for optimal control and indeed specific cases have been found where memory effects are instead detrimental in the presence of control \cite{DDPaper},\cite{Susanna}. 

In this article, we expose the difficulties in employing coherent control to compensate for environment-induced decoherence effects in non-Markovian systems. We consider the widespread assumption (fixed dissipator assumption) that the part of the master equation describing dissipation and dephasing does not change when we add a Hamiltonian control term in the unitary dynamics part. This assumption does not change the physicality of the solutions of the master equation in the Markovian case. We show, however, that this is generally not the case for non-Markovian dynamics. Hence the typical theoretical approaches to quantum control theory cannot be used in the framework of non-Markovian open quantum systems, and only a full microscopic derivation leads to physically meaningful results.

%


For the sake of concreteness we focus on a novel concept utilizing Hamiltonian control recently introduced to counteract the detrimental effect of decoherence. In Ref. \cite{Clemens}, the goal is to seek the  control Hamiltonian that, on asymptotic time scales, optimally upholds a given target property (e.g.~coherence, entanglement or fidelity with respect to a target state). The space of Hamiltonians cannot be efficiently parametrized, hence the problem is approached from a different perspective. The key idea is to optimize some target property in the set of stabilizable cycles, comprising all closed periodic trajectories $\rho(t)=\rho(t+T)$ for which a periodic control Hamiltonian exists such that $\rho(t)$ solves the master equation ($\hbar = 1$)
\eq
\dot\rho=-i[H(t),\rho]+\mathcal{D}(\rho),
\label{ME}
\eeq
with a fixed dissipator $\mathcal{D}(\rho)=\sum_k \gamma_k[L_k\rho L_k^\dagger-\frac{1}{2}\{L_k^{\dagger}L_k,\rho\}]$ composed of Lindblad operators $L_k$ and decay rates $\gamma_k$. The crucial insight behind this is that physically admissible trajectories in state space are strongly constrained by the dissipative part $\mathcal{D}$ of the dynamics alone, which is assumed to be fixed (i.e., it does not change in the presence or absence of the control Hamiltonian). 
A simple criterion, for which a time-dependent control Hamiltonian $H(t)$ exists such that $\rho(t)$ solves the master equation, can be written as follows:
\eq
\forall t \:\forall n:\text{Tr}[\rho^{n-1}(t)\mathcal{D}(\rho(t))]=\frac{1}{n}\delta_t\text{Tr}[\rho(t)^n],
\label{GenF}
\eeq
which holds for any $\rho(t)$ with non-degenerate eigenvalues and $n\in\{2,..,d\}$.  This method thus gives an astonishingly simple characterization of optimal control schemes in the presence of fixed dissipation. Now the obvious questions arise: Can such method be extended to non-Markovian quantum dynamics? Is it possible to determine a time-dependent control Hamiltonian for a fixed non-Markovian dissipator (with temporarily negative decay rates) to optimize some target property? 

In order to tackle these questions let us concentrate on a simple example of a single qubit in a dephasing environment. We begin by exploring the phenomenological case, where the qubit dissipator is fixed under the control sequence. In view of the non-Markovian dynamics considered below, we seek a control Hamiltonian that, on average, optimally upholds the coherence between the ground and excited state for the time $T$ the system markedly evolves (in contrast to asymptotic time scales, as in \cite{Clemens}). Control is no longer required for $t\geq T$; when the Hamiltonian vanishes, the states then cease
to evolve. For the single qubit, an admissible trajectory $\{\rho(t):t\in[0,T)\}$, or $\{\vec r_t\}$ in the Bloch ball, must then satisfy at all times the following criterion \cite{Clemens}:
\eq
\underbrace{\vec r_t.(D\vec r_t+\vec d)}_{f(\vec r_t)}=\underbrace{\frac{1}{2}\delta_t|\vec r_t|^2}_{\dot{p}(\vec r_t)}.
\label{C}
\eeq
Here, the $3\times 3$ matrix $(D)_{ij}=\text{Tr}[\sigma_i\mathcal{D}(\sigma_j)]$ and the vector $(\vec{d})_i=\text{Tr}[\sigma_i\mathcal{D}(\mathbb{I})]$ characterize the dissipator in Bloch notation. Hence, the time evolution of the purity of the state $p=\text{Tr}[\rho^2]=(|\vec r|^2+1)/2$ is exclusively governed by the dissipator. Moreover, for trajectories with purity-increasing sections, one can show that the respective purity flux $f(\rho)=\text{Tr}[\rho\mathcal{D}(\rho)]$ of the dissipator must then be positive. Now, for a Markovian dephasing process, for which the dynamics is generated by the dissipator
\eq
\mathcal{D}_t(\rho)=\frac{\gamma(t)}{2}(\sigma_z\rho\sigma_z-\rho), \label{dedis}
\eeq
with $\gamma(t)>0$, we obtain at all times negative purity flux $f(\vec r, t)=-\gamma(t)(r\sin\phi)^2<0$, where $r\in\{0,1\}$ and $\phi\in\{0,\pi\}$ is the polar angle. Hence, trajectories with purity-increasing sections are physically inconsistent in such system. As a consequence, any trajectory will, after sufficient evolution, inevitably be devoid of coherence, irrespective of any conceivable coherent control strategy. In the presence of non-Markovian effects, however, the decay rate in the dissipator can take negative values, giving rise to periods of negative purity flux. Thus, one may naively think that optimal control trajectories may now be implemented anywhere in the Bloch sphere.

\begin{figure}[htp]
\centering
\includegraphics[width=0.51\textwidth]{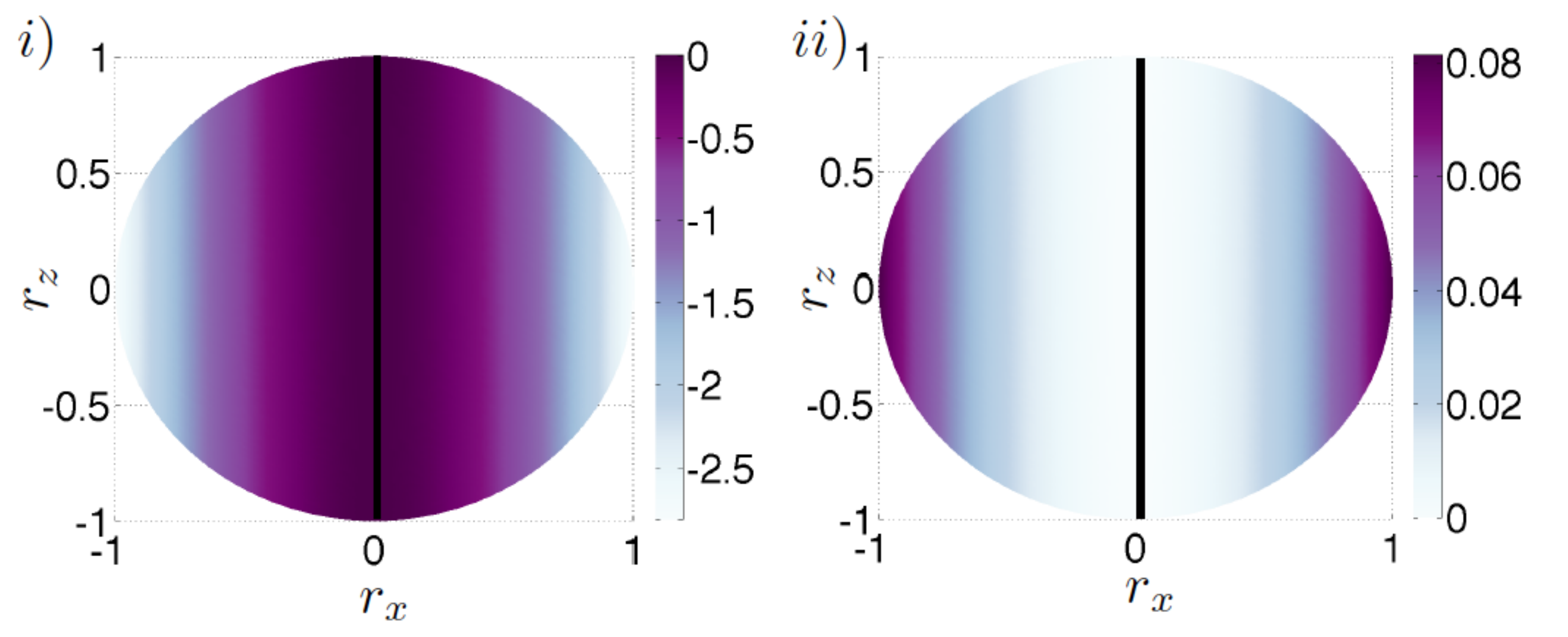}
\caption{Snapshots of the purity flux $f(\vec{r})$ for the purely dephasing dynamics (\ref{dedis}) with the non-Markovian rate (\ref{pdecay}) for $s=3$ and time instances corresponding to positive (i) and negative (ii) values of the decay rate. The set of stabilzable states $\mathcal{S}$ \cite{OCStart} corresponding to vanishing purity flux is time-independent and forms a line along $r_x=0$ (shown in black). For  $\gamma(t)>0$, it is immediate to see that $f(\vec r,t)<0$ (i). On the contrary, $f(\vec r,t)>0$ whenever $\gamma(t)<0$ (ii).}
\label{p1}
\end{figure}

Let us assume that there would exist an implementation strategy, where we can choose a non-Markovian dephasing process which is fixed and independent of the unitary rotations imposed by the control Hamiltonian. We consider a decay rate of the form
\eq
\gamma(t)=[1+t^2]^{-s/2}\Gamma[s]\sin[s\arctan(t)],
\label{pdecay}
\eeq
which is obtained in the exact model of a qubit interacting with a bosonic zero temperature environment with an Ohmic-like spectral density \cite{ohmic1}-\cite{ohmic3},
\eq
J(\omega)=\frac{\omega^s}{\omega_c^{s-1}}e^{-\omega/\omega_c},
\label{spectrum}
\eeq
where $s$ is the Ohmicity parameter and $\omega_c$ a cutoff frequency. The form of spectral density can be modified through the parameter $s$ (the Ohmicity parameter). Specifically, for $s>2$, the decay rate takes temporarily negative values for certain time intervals \cite{Haikka} which temporarily reverses the direction of the purity flux $f(\vec r, t)$ (see Fig.~\ref{p1}). The purity flux associated to this model is shown in Fig.~\ref{p1} for intervals of time when the decay rate is positive (i) and negative (ii) respectively and choosing s = 3 for illustrative purposes.

For the purpose of our optimization task, it is sufficient to restrict to trajectories containing two periods of free evolution (evolution without control), interrupted by a single unitary rotation at the instant $\tilde t$ at which the decay rate changes sign (this is in close analogy to the two-point cycles considered in \cite{Clemens}). The rotation must be fast compared to the incoherent dynamics so that no purity is lost along the way, i.e.~we assume that the rotation is instantaneous. Moreover, we perform the fixed-dissipator assumption, namely we assume that for $t>\tilde t$, the dynamics can still be described with the original Lindbladian $\mathcal{L}_t$, shown in Eq. (\ref{dedis}). For simplicity, we restrict to the case where the rate $\gamma(t)$ changes sign only a single time, i.e.~there exists only one intermediate time $\tilde t>0$ for which $\gamma(\tilde t)=0$ \cite{note}. Trajectories then undergo an initial period of positive decay rate, for which $f(\vec r, t)<0$, followed by a single time period in which the decay rate is negative, and hence $f(\vec r, t)>0$.

\begin{figure}[htp]
\centering
\includegraphics[width=0.52\textwidth]{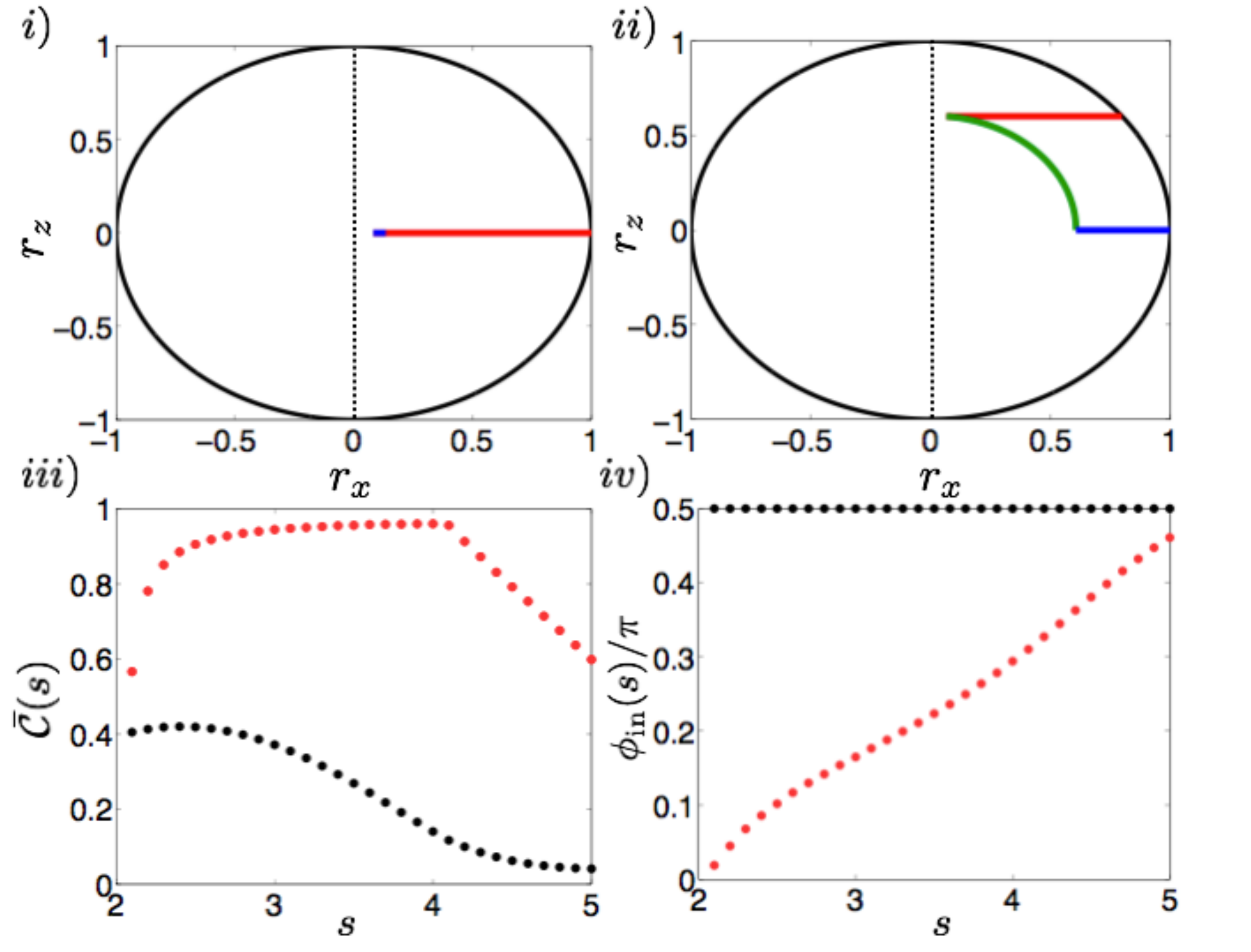}
\caption{Comparison of the optimal trajectories without i) and with ii) coherent control pulses. In the upper panel, we illustrate, for $s=4$, the trajectories in the stages $t< \tilde t$ (red line) and $t>\tilde t$ (blue line), where $\gamma(\tilde t)=0$. The unitary rotation in the controlled trajectory is shown in green. In the lower panel, we show the optimal average coherence $\mathcal{\bar{C}}(s)$ (iii) and corresponding polar angle $\phi_\text{in}(s)$ for the initial Bloch vectors, $r_x(0),\:r_y(0)=r_y(t)=0,\:r_z(0)$ (iv) for the uncontrolled (black) and controlled (red) evolution. For $2<s\leq 4$, we consider a finite dephasing interval $0\leq\omega_c t\leq 30$ \cite{note}. For $4<s\leq 5$, the decay is naturally bound by a time $T$ determined by the Ohmic parameter $s$ \cite{note}. One can clearly see the coherence enhancement obtained in the controlled case.} 
\label{p2}
\end{figure}

In order to indicate the seeming drastic improvement that can be achieved with appropriate coherent control pulses, we compare in Fig.~\ref{p2} for various choices of $s$ the average coherence $\bar{\mathcal{C}}(s)$ that is obtained with the optimal control protocol to the optimal average coherence that is achieved in the uncontrolled case of vanishing Hamiltonian.

The initial conditions are as follows: $r_x(0)=\sin\phi_\text{in}(s),\:r_y(0)=r_y(t)=0,\: r_z(0)=\cos\phi_\text{in}(s)$ where the evolution of $r_x(t)$ and $r_z(t)$ is determined by the dynamics given in Eq. (4). While the respective initial polar angle $\phi_\text{in}(s)$ varies in the controlled case, in the uncontrolled case optimal initial states always lie on the equator of the Bloch ball, corresponding to the region of strongest flux. The physical intuition behind the control protocol is to initially ``kick" the state into the region of weak flux as long as it is exposed to detrimental purity-decreasing dynamics, and to return it to the equator region of maximal flux when it undergoes supportive purity-increasing dynamics. We can regard the initial ``kick" as the equivalent process of preparing a different initial state. It is immediate to see from Fig. 2 iii), that the optimal controlled trajectories are superior in obtaining higher average coherence values as compared to the optimal uncontrolled evolution. In the Ohmic range $2<s\leq 4$, we consider optimal trajectories confined to a time interval $0<t<T>\tilde{t}$. We note that for longer times $T$, the values of optimal average coherence increase but retain the same dependence on the Ohmic parameter $s$. We choose the intermediate rotation to optimally compensate for the weak non-Markovian revivals, in the sense that the purity lost in the red stage is fully regained in the blue stage.
Moreover, we choose to always start the cycle from the surface of the Bloch sphere ($r(0)=1$).  The optimal average coherence and corresponding initial angle $\phi_\text{in}(s)$ (fixed by the constraint $|r(0)|=|r(T)|=1$) increase with $s$ as the strength of the purity flux increases and $\tilde t$ decreases. For $4<s\leq 6$, the initial angle $\phi_\text{in}$ continues to increase but the average coherence $\bar{\mathcal{C}}(s)$ decreases (see Fig. \ref{p2} iii, iv)). For $t \geq T$, the state remains at the surface of the Bloch sphere with perfect coherence. Hence, for asymptotic timescales, coherence can be maximally exploited as a resource. 

These results clearly show the ability to achieve high values of average coherence for non-Markovian dynamics and hence they indicate that, if the fixed-dissipator assumption is satisfied, coherent control schemes could potentially work effectively in the presence of non-Markovian noise. In the specific example considered here, this is even more remarkable since the same method does not work in the Markovian case. On the other hand, for incoherent processes where the method does work in the Markovian case, a wider portion of the state space becomes available for trajectories in the presence of memory effects.

Unfortunately, there exist scenarios where the promise of utilizing non-Markovian dynamics to achieve such high average coherence values can never be achieved experimentally. To show this, let us examine the physical feasibility of the above strategy by critically examining the fixed dissipator assumption in full generality.

Let us consider a fixed dissipator $\mathcal{D}_t$ generating, in the absence of any coherent control, a $t$-parametrized family of completely positive and trace preserving (CPTP) maps $\{\Phi_t\}$, such that $\rho(t)=\Phi_t \rho(0) $. If the dynamics is non-Markovian then the so called divisibility property of the map is not valid. Explicitly, non-divisibility means that the propagator $\Phi_{t,s}=\Phi_t \Phi_s^{-1}$, defined via the relation  $\Phi_t=\Phi_{t,s}\Phi_s$, is not completely positive. Since $\Phi_t$ is, however, CPTP, one should conclude that on a restricted space of initial states (space of accessible states) defined by $\rho(s)=\Phi_s \rho(0)$ the intermediate map $\Phi_{t,s}$ is completely positive. 

If we assume that the dissipator remains fixed under a unitary (coherent) interruption $U$ of the dynamics at time $s$, we can write the dynamical map in the controlled case as $\tilde{\Phi}_t=\Phi_{t,s}U\Phi_s=\Phi_t\Phi_s^{-1}U\Phi_s$. Now, unless the original dynamical maps are covariant, i.e., $U \Phi_t=\Phi_t U$, the object $\tilde{\Phi}_t$ is no longer guaranteed to be a CP map because the unitary can move the intermediate state $\rho(s)$ outside the space of accessible states. Losing complete positivity of the dynamical map means that the dynamics is never physical or, stated another way, no physical implementation of this master equation exists. It is worth noticing that, for Markovian and therefore divisible dynamics this problem does not occur because the propagator is always CP and therefore the modified map $\tilde{\Phi}$ in the presence of coherent control unitaries is always CP and therefore physical.
Thus, in general, knowing the open system dynamics in the absence of control does not give enough information to construct a physically meaningful open system dynamics in the presence of control, even if the control field is completely known. While in very simple cases (two-level system) CP conditions are known, verifying CP of the dynamical map becomes a practically untreatable problem as the dimension of the Hilbert space of the open quantum system increases. In general, the full exact dynamics of the system plus the environment needs to be solved, taking into account the control field in the microscopic derivation, in order to tackle the problem of optimal control in the non-Markovian case. 

To illustrate our result we go back to the simple exact pure dephasing example previously considered and solve the full system plus environment microscopic model in the presence of an instantaneous rotation \cite{ohmic2}.  The only assumption that we will make is that the pulse which induces the unitary rotation is instantaneous. This is a satisfactory approximation provided that the time necessary to perform the pulse is much shorter than any other time scale relevant to the system and is also widely used in descriptions of dynamical decoupled dynamics \cite{dd1, dd2}.
Our aim is to see how the presence of this rotation alters the form of the dissipator and to compare the correct microscopically derived dynamics with the phenomenological approach using the fixed dissipator assumption.

In the absence of control the dynamics are described by the dissipator in Eq. (4) with the decay rate given by Eq.~\eqref{pdecay}. The decay rate is related to the decoherence function $\Gamma(t)$, defined by $\rho_{ij}(t)=\rho_{ij}(0)e^{-\Gamma(t)}$ ($i \neq j$), through the relation $\gamma(t)=\frac{d\Gamma(t)}{dt}$. For the Ohmic class of spectral densities here considered the decoherence function takes the form
\eqa
\Gamma(t)&=&\frac{\Gamma[s]}{s-1}[1-(1+t^2)^{-s/2}\nn\\&\times&(\cos(s\arctan(t))+t\sin(s\arctan(t)))]. \label{decohrate}
\eeqa

We now provide the microscopic derivation of the pure dephasing system subject to a single unitary rotation. We study the exact model for pure dephasing dynamics, with the following Hamiltonian describing the local interaction of a qubit and a bosonic reservoir, in units of $\hbar$: 
\eq
H=\omega_0\sigma_z+\sum_k\omega_ka^\dagger_k a_k+\sum_k\sigma_z(g_ka_k+g^*_ka^\dagger_k)
\eeq
where $\omega_0$ is the qubit frequency, $\omega_k$ the frequencies of the reservoir modes, $a_k(a_k^\dagger)$ the annihilation (creation) operators and $g_k$ the coupling constant between each reservoir mode and the qubit. The form of spectral density can be modified through the parameter $s$ (the Ohmicity parameter). The initial state, composed by the qubit and the field is $\ket{\Psi(0)}=(c_e\ket{e}+c_g\ket{g})\otimes\ket{0}$, with $\ket{0}=\bigotimes_k\ket{0}_{\bf k}$. Let us denote the time at which the unitary pulse is applied by $\tilde{t}$. The system evolves for $0<t<\tilde t$ as follows \cite{ohmic2}: 
\eqa
\ket{\Psi(t)}&=&U(t,0)\ket{\Psi(0)}\nn\\&=&c_e\ket{e}\otimes \ket{\Psi_e(t)} +c_g\ket{g}\otimes  \ket{\Psi_g(t)},\nn\\
\eeqa
with $\ket{\Psi_e}=\bigotimes_k U_e(t,0)\ket{0_{\bf k}}$ and $\ket{\Psi_e}=\bigotimes_k U_e(t,0)\ket{0_{\bf k}}$. The time evolution operator (in the interaction picture) takes the form: 
\eqa
U(t)&=&\text{exp}\left\{-i\int_0^t\sum_k\sigma_z(g_kb_k^{\dagger}e^{i\omega_kt'}+g_k^*b_ke^{-i\omega_kt'})dt'\right\}\nn\\&=&\text{exp}\left\{\sigma_z\frac{1}{2}\sum_k(b_k^{\dagger}\xi_k(t)-b_k\xi^*_k(t))\right\},
\eeqa
with 
\eq
\xi_k(t)=2g_k\frac{1-e^{i\omega_kt}}{\omega_k}.
\eeq
Here, $U(t)$ can be described as a conditional displacement operator, the sign of the displacement being dependent on the logical value of the qubit, denoted $U_e$ and $U_g$ for the respective values. In particular, for any pure state $\ket{\Phi}$ of the field: 
\eqa
& &U(t)\ket{g}\otimes\ket{\Phi}=\ket{g}\otimes\prod_kD\left(-\frac{1}{2}\xi_k(t)\right)\ket{\Phi} \nn\\& & 
U(t)\ket{e}\otimes\ket{\Phi}=\ket{e}\otimes\prod_kD\left(\frac{1}{2}\xi_k(t)\right)\ket{\Phi}
\eeqa
where the displacement operator $D(\xi_k)$ is defined as:
\eq
D(\xi_k)=\text{exp}\{b_k^{\dagger}\xi_k-b_k\xi^*_k\}
\eeq
and $\ket{\frac{1}{2}\xi_k(t)}$ is a coherent state of amplitude $\frac{1}{2}\xi_k(t)$. If a rotation occurs at $\tilde t$, we have: 

\eqa
& &\ket{\Psi_\text{ROT}(\tilde t)}=R_y(\phi)\ket{\Psi(\tilde t)}\nn\\&=&c_e\left(\cos\left(\frac{\phi}{2}\right)\ket{e}+\sin\left(\frac{\phi}{2}\right)\ket{g}\right)\otimes U_e(\tilde t,0)\ket{0}\nn\\&+&c_g\left(-\sin\left(\frac{\phi}{2}\right)\ket{e}+\cos\left(\frac{\phi}{2}\right)\ket{g}\right)\otimes U_g(\tilde t,0)\ket{0}\nn\\&=&\ket{e}\otimes[c_e\cos\left(\frac{\phi}{2}\right)U_e(\tilde t,0)\ket{0}-c_g\sin\left(\frac{\phi}{2}\right)U_g(\tilde t,0)\ket{0}]\nn\\&+&\ket{g}\otimes[c_e\sin\left(\frac{\phi}{2}\right)U_e(\tilde t,0)\ket{0}+c_g\cos\left(\frac{\phi}{2}\right)U_g(\tilde t,0)\ket{0}].\nn\\
\eeqa

and $R_y(\phi)=e^{-i\frac{\phi}{2}\sigma_y}=\cos\frac{\phi}{2}\mathbb{I}-i\sin\frac{\phi}{2}\sigma_y$. 
For times $t>\tilde t$, the combined system evolves according to: 
\eqa
& &U(t,\tilde t)R_y(\phi)U(\tilde t,0)\ket{\Psi(0)}\nn\\&=&\ket{e}\otimes[c_e\cos\left(\frac{\phi}{2}\right)\ket{\Psi_{ee}(t)}-c_g\sin\left(\frac{\phi}{2}\right)\ket{\Psi_{eg}(t,\tilde t)}]\nn\\&+&\ket{g}\otimes[c_e\sin\left(\frac{\phi}{2}\right)\ket{\Psi_{ge}(t,\tilde t)}+c_g\cos\left(\frac{\phi}{2}\right)\ket{\Psi_{gg}(t)}]\nn\\& &,
\eeqa
where we have: 
\eqa
& & \ket{\Psi_{ee}(t)}=U_e(t,\tilde t)U_e(\tilde t,0)\ket{0} \nn\\& & 
\ket{\Psi_{ge}(t,\tilde t)}=U_g(t,\tilde t)U_e(\tilde t,0)\ket{0} \nn\\& & 
\ket{\Psi_{eg}(t,\tilde t)}=U_e(t,\tilde t)U_g(\tilde t,0)\ket{0}\nn\\& & 
\ket{\Psi_{gg}(t)}=U_g(t,\tilde t)U_g(\tilde t,0)\ket{0}\nn\\
\eeqa

The matrix elements of the reduced density matrix of the qubit are defined as: 
\eq
\rho_{ij}(t,\tilde t)=\bra{i}\text{Tr}_EU(t,\tilde t)R_y(\Phi)U(\tilde t,0)\rho(0)U^{\dagger}(\tilde t,0)R_y^{\dagger}(\Phi)U^{\dagger}(t,\tilde t)\ket{j}.
\eeq

Hence, we have the following elements of the density matrix, 
\eqa
\rho_{ee}(t,\tilde t)&=&|c_e|^2\cos^2\left(\frac{\phi}{2}\right)+|c_g|^2\sin^2\left(\frac{\phi}{2}\right)\nn\\&-&(c_ec_g^*\langle\Psi_{eg}(t,\tilde t)|\Psi_{ee}(t)\rangle\nn\\&+&c_e^*c_g\langle\Psi_{ee}(t)|\Psi_{eg}(t,\tilde t)\rangle
)\sin\left(\frac{\phi}{2}\right)\cos\left(\frac{\phi}{2}\right)\nn\\
\eeqa
\begin{eqnarray}
\rho_{eg}(t,\tilde t)&=&\cos\left(\frac{\phi}{2}\right)\sin\left(\frac{\phi}{2}\right)(|c_e|^2\langle\Psi_{ee}(t)|\Psi_{ge}(t,\tilde t)\rangle \nn\\&-&|c_g|^2\langle\Psi_{eg}(t,\tilde t)|\Psi_{gg}(t)\rangle)\nn\\&+&c_e^*c_g\cos^2\left(\frac{\phi}{2}\right)\langle\Psi_{ee}(t)|\Psi_{gg}(t)\rangle\nn\\&-&c_ec_g^*\sin^2\left(\frac{\phi}{2}\right)\langle\Psi_{eg}(t,\tilde t)|\Psi_{ge}(t, \tilde t)\rangle.\nn\\
\end{eqnarray}
where $\rho_{gg}=1-\rho_{ee}$ and $\rho_{ge}=\rho_{eg}^*$. 

Before the unitary rotation, the dynamics of the qubit are of course described via the decoherence function of Eq. \eqref{decohrate}.  After an instantaneous rotation of angle $\phi$ around the y-axis, the resultant Bloch vectors are given as follows:
\eqa
r_x(t,\tilde t)&=&r_z(0)\sin(\phi)e^{-\Gamma(t-\tilde t)}\cos[y(t)]+r_x(0)e^{-\Gamma(t)}\nn\\&\times&[\cos^2\left(\frac{\phi}{2}\right)-\sin^2\left(\frac{\phi}{2}\right)e^{2[\Gamma(t)-\Gamma(\tilde t)-\Gamma(t-\tilde t)]}]\nn\\& &
\eeqa
\eqa
r_y(t,\tilde t)&=&r_z(0)\sin(\phi)e^{-\Gamma(t-\tilde t)}\sin[y(t)]+r_y(0)e^{-\Gamma(t)}\nn\\&\times&[\cos^2\left(\frac{\phi}{2}\right)-\sin^2\left(\frac{\phi}{2}\right)e^{2[\Gamma(t)-\Gamma(\tilde t)-\Gamma(t-\tilde t)]}]\nn\\& &
\eeqa
\eqa
r_z(\tilde t)=r_z(0)\cos(\phi)-r_x(0)\sin(\phi)e^{-\Gamma(\tilde t)}
\eeqa
where $y(t)=\text{Im}(\tilde\Gamma(t)-\tilde\Gamma(\tilde t)-\tilde\Gamma(t-\tilde t))$ and
\eqa
\tilde\Gamma(t)&=&4\Gamma[s-1]\nn\\&\times&(1-t^2)^{-s/2}[\sin(s\arctan(t))-t\cos(s\arctan(t)].\nn\\
\eeqa

\begin{figure}[htp]
\centering
\includegraphics[width=0.35\textwidth]{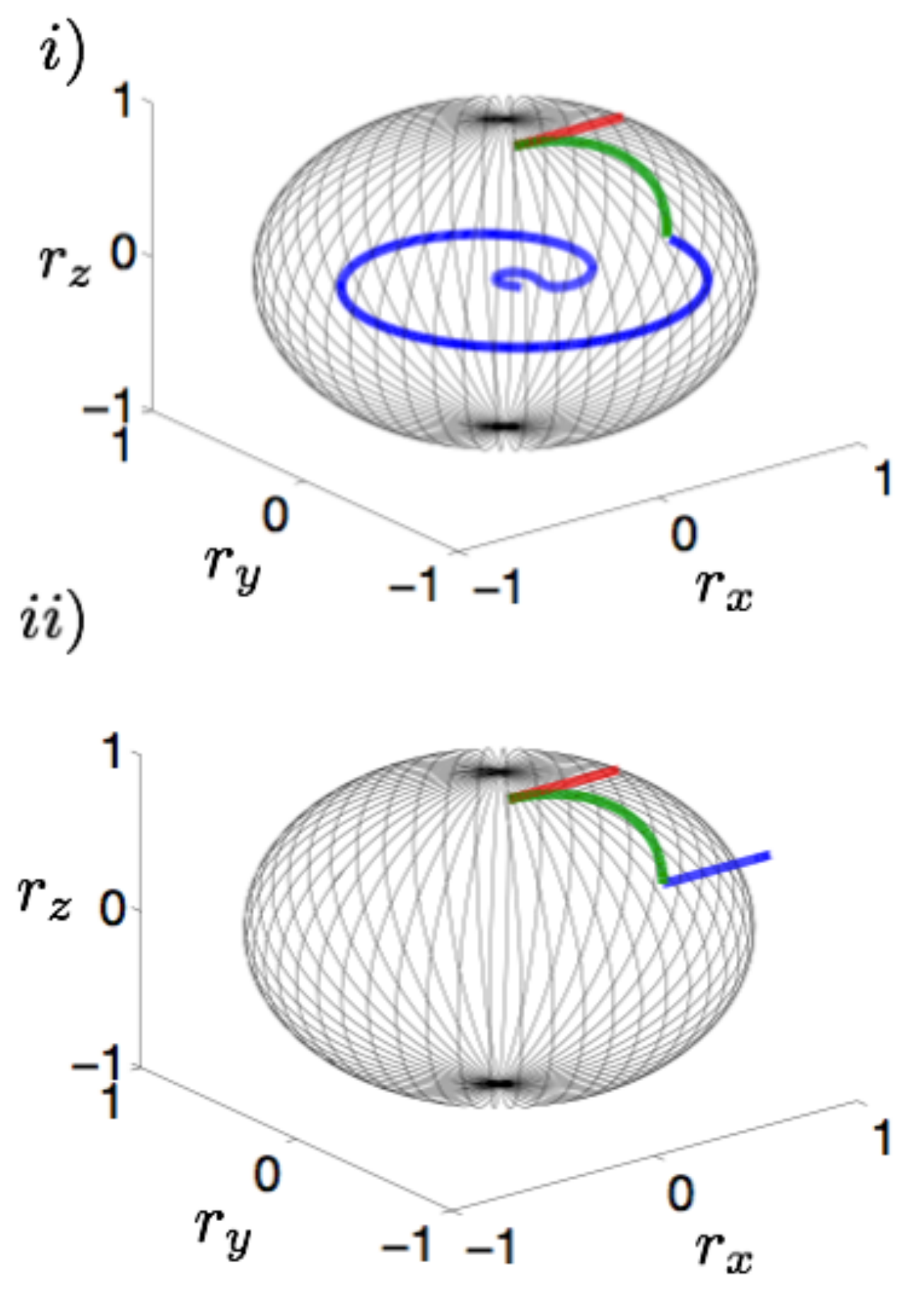}
\caption{Dynamics of the qubit obtained from i) the exact model in the presence of a control pulse and ii) assuming a fixed dissipator. The trajectory is divided into three stages:  $t < \tilde t$ (top red line), $t>\tilde t$ (bottom blue line) and the unitary jump between the two (middle green line). For illustrative purposes, we have chosen $s=4$ and initial polar angle $\Phi_\text{in}(s)>\tilde\Phi_\text{in}=0.2\pi$. }
\label{bloch}
\end{figure}

In Fig.~\ref{bloch} i), we plot the true evolution of the system, when the dissipator is not fixed but instead the unitary rotation is incorporated into the microscopic derivation. 
This dynamics should be contrasted with the corresponding trajectory on the Bloch sphere in the fixed dissipator assumption, as shown in Fig.~\ref{bloch} ii). The latter dynamics are obviously not CP, since the trajectory falls outside the Bloch sphere. Such trajectories can be achieved, for example, by choosing initial aximuthal angles $\tilde\phi_{\text{in}}<\phi_{\text{in}}(s)$, where $\phi_{\text{in}}(s)$ is the angle fixed by the constraint $|r(0)|=|r(T)|=1$, shown in in Fig. 2 iv). 

To conclude, we briefly comment on the special case in which the fixed dissipator assumption is used but the qubit undergoes a covariant dynamics.  
%
It is easy to convince oneself that coherent control will never be useful in such a case, since by definition applying the control pulse before, during, or after the non-unitary evolution leads to the very same state. Hence,  choosing a different initial state is equivalent to implementing any coherent control sequence during the evolution.
 

Summarizing, our results show that the appealing idea of using optimal control strategies in the presence of non-Markovian noise inevitably leads to formidable difficulties. Indeed, except for a few simple cases in which CP of the controlled open dynamics can be checked, the only physically meaningful description of the reduced dynamics in the presence of control pulses currently appears to be the one obtained via an exact microscopic model entailing system plus controlled pulses plus interaction with the environment. If the decay rate in the dissipator changes markedly after each pulse, it is clear that, in general, constructing superior trajectories will never be a feasible task. A possible solution to this impasse might be the discovery of certain special forms of non-Markovian dissipators that may not be changing sensibly in the presence of some specific coherent control schemes, perhaps exploiting specific symmetries


C.A. acknowledges financial support from the EPSRC (UK) via the Doctoral Training Centre in Condensed Matter Physics under Grant no EP/G03673X/1. S.M. and E-M.L acknowledge support by the EU Collaborative project QuProCS (Grant Agreement 641277), the Academy of Finland (Project no. 287750) and the Magnus Ehrnrooth Foundation.


\begin{thebibliography}{30}
\bibitem{note} For this reason, we consider the Ohmic range $2<s\leq 5$ where information backflow occurs in the time interval $t\in[t_1,t_2]$ and $t_1=\tan(\pi/s)$ and $t_2=\infty(\tan(2\pi/s)$ for $2<s\leq 4$ and $4<s\leq 5$ respectively. Decay rates with $n$ intervals of information backflow (for $s>6$) can be implemented as $n$ cycles joint through a series of unitary transformations.
\end{thebibliography}
\end{document}